\documentclass[twocolumn,letter]{emulateapj}

\usepackage{graphicx}
\usepackage{latexsym,amssymb}
\usepackage{amsmath}
\usepackage{natbib}
\newcommand{\kms}{km\,s$^{-1}$}
\newcommand{\dgr}{$^\circ$}
\newcommand{\Msun}{M$_\odot$}
\newcommand{\Msunyr}{M$_\odot$\,yr$^{-1}$}
\newcommand{\kmskpc}{km\,s$^{-1}$\,kpc$^{-1}$}
\newcommand{\ccm}{cm$^{-3}$}
\newcommand{\rmd}{\mathrm{d}}

\newcommand{\HI}{H{\sc i}}
\newcommand{\NII}{[N{\sc ii}]}

\newcommand{\SIIdoublet}{\hbox{[S{\sc ii}]$\lambda\lambda$6716.4,6730.8}}
\newcommand{\hst}{\texttt{HST}}
\newcommand{\vlt}{\texttt{VLT}}
\newcommand{\gifu}{GMOS-IFU}

\newcommand{\Vs}{V_\star}
\newcommand{\RCR}{R_\mathrm{CR}}
\newcommand{\avne}{n_e}
\newcommand{\dne}{\Delta n_e}
\newcommand{\drgas}{\Delta \rho_\textrm{gas}}
\newcommand{\Mbh}{M_\textrm{BH}}
\newcommand{\Mdot}{\dot{M}}
\newcommand{\MEdd}{\dot{M}_\textrm{Edd}}

\shorttitle{Kinematic analysis of nuclear spirals} 
\shortauthors{van de Ven \& Fathi}

\begin{document}

\title{Kinematic analysis of nuclear spirals: feeding the black hole
  in NGC\,1097}

\author{Glenn van de Ven\altaffilmark{1,2}, Kambiz
  Fathi\altaffilmark{3,4}}

\altaffiltext{1}{Max Planck Institute for Astronomy, K\"onigstuhl 17,
  69117 Heidelberg, Germany: glenn@mpia.de}

\altaffiltext{2}{Institute for Advanced Study, Einstein Drive, Princeton,
  NJ 08540, USA; Hubble Fellow}

\altaffiltext{3}{Stockholm Observatory, Department of Astronomy,
  Stockholm University, AlbaNova, 106 91 Stockholm, Sweden;
  kambiz@astro.su.se}

\altaffiltext{4}{Oskar Klein Centre for Cosmoparticle Physics, Stockholm
  University, 106 91 Stockholm, Sweden}

\begin{abstract}
  We present a harmonic expansion of the observed line-of-sight velocity
  field as a method to recover and investigate spiral structures in
  the nuclear regions of galaxies. 
  We apply it to the emission-line velocity field within the
  circumnuclear star forming ring of NGC\,1097, obtained with the
  GMOS-IFU spectrograph.
  The radial variation of the third harmonic terms are well described
  by a logarithmic spiral, from which we interpret that the
  gravitational potential is weakly perturbed by a two-arm spiral
  density wave with inferred pitch angle of $52 \pm 4$\dgr. This
  interpretation predicts a two-arm spiral distortion in the surface
  brightness, as hinted by the dust structures in central images of
  NGC\~1097, and predicts a combined one-arm and three-arm spiral
  structure in the velocity field, as revealed in the non-circular
  motions of the ionised gas.
  Next, we use a simple spiral perturbation model to constrain the
  fraction of the measured non-circular motions that is due to radial
  inflow. We combine the resulting inflow velocity with the gas
  density in the spiral arms, inferred from emission line ratios, to
  estimate the mass inflow rate as a function of radius, which reaches
  about $0.011$\,\Msunyr\ at a distance of $70$\,pc from the center.
  This value corresponds to a fraction of about $4.2 \times 10^{-3}$
  of the Eddington mass accretion rate onto the central black hole in
  this LINER/Seyfert1 galaxy.
  We conclude that the line-of-sight velocity not only can provide a
  cleaner view of nuclear spirals than the associated dust, but that
  the presented method also allows the quantitative study of these
  possibly important links in fueling the centers of galaxies,
  including providing a handle on the mass inflow rate as a function
  of radius.
\end{abstract}


\section{Introduction}
\label{sec:intro}

Gas transport to the centers of galaxies is still mainly an unsolved
problem \citep[e.g.][]{Martini2004}. Since most of this gas is
residing in a rotating disk well beyond the center, it is essentially a
problem of angular momentum transport.
Proposed transport mechanisms range from galactic interactions and
bars, to nuclear bars and spirals, to stellar mass loss and disruption
near the central black hole (BH). This range in scales also closely
represents a range in decreasing mass inflow rates, which in turn
might be correlated with activity in the galactic nucleus (AGN),
ranging from quasar, Seyfert, LINER to quiescent galaxies.
However, not only is it challenging to (observationally) establish
fueling mechanisms down to a few parsec from the center, time delays
between changes in the mass inflow rate and the onset of nuclear
activity further complicate linking them.

It is also likely that multiple fueling mechanisms are important and
act together.
Large-scale bars are efficient at transporting gas inward
\citep[e.g.][]{Athanassoula1992}, but the presence of an inner
Lindblad resonance (ILR) will cause the gas to pile up in a nuclear
ring, often clearly visible due to the intense star formation (see
e.g.\ Figure~\ref{fig:images}).
The gas might continue further inward through nested bars
\citep[e.g.][]{Shlosman1989, Englmaier2004}, although dynamical
constraints on a double-barred system may prohibit inflow down to the
center \citep{Maciejewski2002}. Moreover, observational evidence of
nuclear bars is scarce \citep[e.g.][]{Regan1999,Martini1999}, but
should be interpreted with care since the prominent dust lanes in the
main bar might be absent in secondary bars
\citep[e.g.][]{Shlosman2002}.

On the other hand, nuclear spirals seem to be commonly observed, both
in active and quiescent galaxies \citep[e.g.][]{Laine1999,Pogge2002},
ranging from flocculent to grand-design nuclear spirals
\citep[e.g.][]{Martini2003}.
Whereas the former are suggested to form by acoustic instabilities
\citep[e.g.][]{Elmegreen1998}, the grand-design nuclear spirals are
thought to be the result of gas density waves
\citep[e.g.][]{Englmaier2000} or shocks
\citep[e.g.][]{Maciejewski2004a, Maciejewski2004b} induced by the
non-axisymmetric gravitational potential of a large-scale bar.
The latter two might be connected in the sense that the bar-driven
spiral shocks trigger the gas density waves throughout the disk
\citep[][]{Ann2005}, which in turn seem to be necessary for the
nuclear spiral to be long-lived \citep{Englmaier2000}.
The inward extended gas inflow through a nuclear spiral not only
depends on the torque of the large-scale bar, but also the gas having
high enough sound speed so as to loose angular momentum
\citep[e.g.][]{Englmaier1997, Patsis2000}, as well as the presence of
a central mass concentration such as a super-massive BH to overcome a
closer-in (inner) ILR \citep[e.g.][]{Fukuda2000, Ann2005}.

The deviations in the gas density due to nuclear spirals are typically
only a few per cent \citep[][]{Englmaier2000}, which makes direct
imaging very difficult. Instead, the obscuration due to dust thought
to be associated with the gas overdensities is often employed, but
when the extinction is small or the dust not well-mixed with the gas,
it can lead to unclear or even missed detections of nuclear spirals.
At the same time, nuclear spirals induce non-circular motions in the
gas, with resulting deviations in the observed velocity field that can
be a significant fraction of the underlying circular velocity. If the
gas is ionized, the kinematics might be inferred from emission lines,
which in general can be detected more easily and at higher spatial
resolution in the optical, than can be tracers of the molecular or
atomic gas such as CO and \HI\ at radio wavelengths. Moreover, while
imaging typically only yields a detection of the nuclear spiral, the
observed non-circular motions in combination with the gas
(over)density inferred from the simultaneously measured (line) fluxes
might be used to derive an estimate of the gas mass in/out-flow rate.

An alternative approach to get a handle on the gas flow rates is to
compute the gravitational torques from the observed surface brightness
using a mass-to-light ratio conversion calibrated against (circular)
velocity measurements \citep{GarciaBurillo2005}, or against stellar
population models fitted to color measurements \citep{Quillen1995a}.
In these cases, various assumptions are made, most importantly that
stellar light is a clean (once corrected for dust obscuration) and
direct (mass follows light) tracer of the underlying gravitational
potential.  Even so, this approach shows that gravitational torques
due to non-axisymmetric structures are very efficient at transporting
gas inward while overcoming dynamical barriers such as the corotation
resonance.
However, when the resulting predicted mass inflow rates as functions
of radius are compared with the non-circular motions in the observed
gas velocity fields the correlation between both is often not evident
\citep{Haan2009}.

The main difficulty with the non-circular motions is to identify the
fraction that is due to pure radial flow. Gas on closed elliptic
orbits not only contributes to non-circular motions (or elliptic
streaming) in the azimuthal direction but also in the radial direction
\citep{Wong2004}. However, since closed elliptic orbits are applicable
strictly only to collisionless (stellar) orbits, we expect that as a
result of shocks and other dissipational effects the angular momentum
of the gas will change, leading to in/out-flow. While the measured
non-circular motions are often taken to be directly representative of
the radial flow velocities \citep[e.g.][]{StorchiBergmann2007a}, only
a fraction of them are expected to be truly radial in/out-flows.

The goal of this paper is two-fold: (i) to show that harmonic
expansion of gas velocity fields provides a clean way to detect
nuclear spirals and to estimate their pitch angles; (ii) to use simple
perturbation models to constrain the fraction of measured non-circular
motions that is due to radial flow, and to estimate the corresponding
mass inflow rate.
In Section~\ref{sec:harmanalysis} we present the harmonic analysis,
and in Section~\ref{sec:ngc1097} we apply it to the observed
emission-line velocity field within the circumnuclear star forming ring
of NGC\,1097. We fit the third harmonic terms as a three-arm
logarithmic spiral structure consistent with a weak two-arm spiral
perturbation of the gravitational potential. We then use the
perturbation model derived in Appendix~\ref{app:gasorbits} to
constrain the inflow rate as function of distance from the center of
NGC\,1097. In Section~\ref{sec:discussion} we discuss the
corresponding spiral distortion in the surface brightness, and
possible additional non-circular motion contributions to the velocity
field. Finally, we link the estimated mass inflow rate to the
accretion onto the central BH of NGC\,109. We summarize and draw our
conclusions in Section~\ref{sec:summaryandconcl}.

\section{Harmonic analysis}
\label{sec:harmanalysis}

Maps of the surface brightness (SB), line-of-sight velocity ($V$),
velocity dispersion ($\sigma$), and higher order velocity moments
(often expressed in term of Gauss-Hermite moments $h_3$, $h_4$, \dots)
of nearby galaxies generally show organized, periodic features, which
can be suitably studied by means of a harmonic expansion
\citep[e.g.][]{Schoenmakers1997, Wong2004, Fathi2005, Krajnovic2006,
  Spekkens2007}.

\subsection{Harmonic expansion}
\label{sec:expansion}

We start by dividing a map into a number of elliptic annuli, each with
a different semi-major axis length $R$, but we assume all have the
same flattening $q$ and share the same angle $\psi_0$ and center
$(x'_0,y'_0)$, so that in terms of Cartesian coordinates on the map
(i.e., on the plane of the sky)
\begin{eqnarray}
  \label{eq:ellannuli}
  x' = x'_0 + R \cos\psi \cos\psi_0 - q\,R \sin\psi \sin\psi_0,
  \nonumber \\
  y' = y'_0 + R \cos\psi \sin\psi_0 + q\,R \sin\psi \cos\psi_0.
\end{eqnarray}
If the map is aligned such that $x'$ is pointing North and $y'$
pointing West, than $\psi_0=\pi/2+\Gamma$, where $\Gamma$ is the
common (observational) definition of the position angle of the major
axis of the galaxy, measured from North through East. Next, we extract
the profiles along each of the annuli and describe them by a finite
number of harmonic terms $n$,
\begin{equation}
  \label{eq:harmdefAB}
  P = c_0(R) + 
  \sum_{m=1}^n c_m(R) \cos m\psi + s_m(R) \sin m\psi.
\end{equation}
There are various ways to obtain the set of best-fit ellipses, but
given the highly non-linear nature of the optimization for the
best-fit parameters, this is commonly done in a stepwise, iterative
process. We follow a similar approach as described in
\cite{Krajnovic2006} --- based on the usually adapted procedure in
photometry \citep{Jedrzejewski1987} --- which we briefly summarize.

For (early-type) galaxies, the maps of SB, $\sigma$ and \textit{even}
higher-order velocity moments ($h_4$, $h_6$, \dots) each are to first
order well approximated by a function that is constant along
(concentric, equally flattened and orientated) ellipses, and only
varies as function of the semi-major axis length of these ellipses.
Similarly, $V$ and \textit{odd} higher-order velocity moments ($h_3$,
$h_5$, \dots) each are well described by a similar function, but one
that has an additional cosine variation along the ellipses. In these
cases, the harmonic expansion in equation~\eqref{eq:harmdefAB}
truncates after the third-order term ($n=3$). The parameters of the
best-sampling ellipses for even velocity moments, can then be obtained
by minimizing $\chi^2 = \Sigma_{m=1}^3 (c_m^2 + s_m^2)$, and without
$c_1$ for odd velocity moments. We first perform the minimization for
a grid of (fixed) $q$ and $\psi_0$ values, where we might use external
constraints such as the inclination $i$ and measured position angle
$\Gamma$ together with an initial estimate of the center
$(x'_0,y'_0)$. Next, starting from the best-fit grid pair, we optimize
for all four parameters.

We sample the semi-major axis lengths as $R = R_1 [ k + (1+g)^{(k-1)}
]$ for $k=1,2,3,\dots$, with the initial $R_1$ depending on the
spatial resolution; for the geometric increase factor we take $g=0.1$.
We achieve this sampling, together with uniform sampling in $\psi$, by
bilinear interpolation of the observed map. When fitting the observed
line-of-sight velocity field shown in Figure~\ref{fig:hardec_maps}, we
assume all motions are within the equatorial plane. We divide all
harmonic terms by $q=\sin i$ to take into account the projection
effect. For further details, including error estimates, see
\cite{Fathi2005}.

\subsection{Spiral structure}
\label{sec:spiralstructure}

Once we have obtained the set of best-fitting ellipses, we fit the
harmonic expansion of equation~\eqref{eq:harmdefAB} to each of the
corresponding profiles to obtain the harmonic terms $c_m$ and $s_m$
(up to higher order $n>3$), as functions of $R$. The difference with
the above expansion up to and including $n=3$, reveals the deviations
from the latter smooth model: for example, boxiness and diskiness in
the SB map, and non-circular motion in the residual $V$ map.

In the case of a spiral structure as observed in NGC\,1097, both
through the dust in the SB and in the residual $V$ map
\citep{Fathi2006}, the deviations are more naturally described as a
(radially varying) offset in the angle $\psi$, than through a complex
combination of variations in the amplitudes $c_m$ and $s_m$.
Henceforth, we rewrite the harmonic expansion of
equation~\eqref{eq:harmdefAB} (in a mathematically equivalent
expression) as
\begin{equation}
  \label{eq:harmdefCph}
  P = K_0(R) + 
  \sum_{m=1}^n K_m(R) \cos\left(m[\psi-\psi_m(R)]\right),
\end{equation}
where the amplitudes $K_m$ and phase shifts $\psi_m$ are related to
the coefficients $c_m$ and $s_m$ ($s_0=0$) by
\begin{equation}
  \label{eq:relABandCph}
  K_m^2 = c_m^2 + s_m^2
  \quad \mathrm{and} \quad
  \tan(m\psi_m) = \frac{s_m}{c_m}.
\end{equation}
For example, the $V$ map of NGC\,1097 in Figure~\ref{fig:hardec_maps}
reveals, after subtracting the best-fit circular motion $K_0 + K_1(R)
\cos\psi$ (with $K_1=c_1$, since we set $\psi_1=s_1=0$), what seems to
be a three-arm spiral structure, $K_3(R) \cos(3[\psi-\psi_3(R)])$.
Here, $\psi_3(R)$ traces the spiral arms as function of radius $R$.

\subsection{Weakly perturbed gravitational potential}
\label{sec:potentialpertubation}

An axisymmetric gravitational potential in a frame that co-rotates with
a weak perturbation of harmonic number $m$ can be written in polar
coordinates as
\begin{equation}
  \label{eq:potharmm}
  \Phi(R,\phi) = \Phi_0(R) + \Phi_m(R)
  \cos\left(m\left[\phi - \phi_m(R)\right]\right),
\end{equation}
where $\phi_m$ is the phase of the perturbation. Through Poisson's
equation it follows that the corresponding surface mass density
exhibits a harmonic $m$ distortion. 

To derive the line-of-sight velocity, we follow
\cite{Schoenmakers1997}. As described in Appendix~\ref{app:gasorbits},
we extend their collisionless analysis by including radial damping in
the equations of motion to take into account the dissipative nature of
gas. We assume that the gas moves on closed loop orbits in the
equatorial plane, which we observe at an inclination $i$ away from its
normal and at an (azimuthal) angle $\phi_\mathrm{los}$. Given a point
$(R,\psi)$ in the equatorial plane, the projection of the azimuthal
and radial velocity onto the corresponding line-of-sight yields
\begin{equation}
  \label{eq:vlosharm}
   V_\mathrm{los} = \sin i \left[ v_\phi(R,\psi) \cos\psi 
     + v_R(R,\psi) \sin\psi \right],
\end{equation}
where $\psi=\phi-\phi_\mathrm{los}+\pi/2$ is zero on the line of nodes
\citep[see also Fig.~1 of][]{Schoenmakers1997}. To first order the
solutions of the equations of motion in the perturbed gravitational
potential of equation~\eqref{eq:potharmm} yield
\begin{eqnarray}
  \label{eq:solvR}
  v_R(R,\psi) & = & v_c(R) 
  \left[ c_R \cos m\psi + s_R \sin m\psi \right], 
  \\
  \label{eq:solvphi}
  v_\phi(R,\psi) & = & v_c(R)
  \left[ 1 + c_\phi \cos m\psi + s_\phi \sin m\psi\right],
\end{eqnarray}
where $v_c^2 = R \rmd \Phi_0/\rmd R$ is the circular velocity and
$c_R$, $s_R$, $c_\phi$ and $s_\phi$ are functions of $R$ given in
Appendix~\ref{app:gasorbits}. Substituting these solutions into
equation~\eqref{eq:vlosharm}, we obtain
\begin{eqnarray}
  \label{eq:solvlos}
  V_\mathrm{los} & = & \Vs \cos\psi
    \nonumber \\
    & + & c_{m-1} \cos(m-1)\psi + s_{m-1} \sin(m-1)\psi
    \nonumber \\
    & + & c_{m+1} \cos(m+1)\psi + s_{m+1} \sin(m+1)\psi, 
\end{eqnarray}
with $c_{m\pm1} = \Vs (c_\phi \mp s_R)/2$ and $s_{m\pm1} = \Vs (s_\phi
\pm c_R)/2$, and $V_\star \equiv v_c(R) \sin i$ the circular velocity
in projection. We thus find, as concluded before by
\cite{Schoenmakers1997} and already qualitatively inferred by
\cite{Canzian1993} that, if the gravitational potential has a
perturbation of harmonic number $m$, the line-of-sight velocity field
contains an $m-1$ and an $m+1$ harmonic term.

\subsection{Pitch angle}
\label{sec:pitchangle}

How loosely or tightly wound a spiral is can be quantified via its
pitch angle, $\zeta$, which, at a given radius, $R$, measures the
angle between the tangent of the spiral arm and a circle with radius
$R$ in the plane of the disk. Inverting the phase-shift, $\phi(R)$, of
a spiral, we can parameterize an arm of the spiral as
\begin{equation}
  \label{eq:partzmspiral}
  x = R(\phi)\,\cos\phi
  \quad \mathrm{and} \quad
  y = R(\phi)\,\sin\phi.
\end{equation}
One can then show that the pitch angle $\zeta$ is given by
\begin{equation}
  \label{eq:pitchangle}
  \cot\zeta = \frac{\rmd \phi}{\rmd \ln R},
\end{equation}
which is positive (negative) if the spiral curves anti-clockwise
(clockwise) with increasing radius. The smaller the pitch angle, the
more tightly the spiral is wound, with $\zeta=0$ a circle, while
$\zeta=\pm\pi/2$ corresponds to a straight line. A specific case that
is often encountered in nature is that of a logarithmic spiral
\begin{equation}
  \label{eq:logmspiral}
  \phi(R) = \frac{1}{b_0} \ln\frac{R}{a_0}
  \quad \leftrightarrow \quad
  R(\phi) = a_0\,\exp(b_0\,\phi),
\end{equation}
with constants $a_0$ and $b_0$. From equation~\eqref{eq:pitchangle},
we find the well-known property that the logarithmic spiral has a
constant pitch angle $\zeta = \tan^{-1} b_0$.

In the case of an $(m\pm1)$-spiral in the line-of-sight velocity
field, the corresponding pitch angle, $\zeta_{m\pm1}$, follows
directly from the phase-shift, $\psi_{m\pm1}(R)$, in the harmonic
expansion in equation~\eqref{eq:relABandCph}. The pitch angles
$\zeta_{m-1}$ and $\zeta_{m+1}$ generally take different values and
bracket the pitch angle $\zeta_m$ of the $m$-spiral perturbation in
the gravitational potential that caused them. To show this, we start
from equations~\eqref{eq:appsolcmpm1} and~\eqref{eq:appsolsmpm1} in
Appendix~\ref{app:gasorbits} and rewrite the coefficients $c_{m\pm1}$
and $s_{m\pm1}$ as
\begin{eqnarray}
  \label{eq:harmcoeffphim}
  c_{m\pm1} & = & K_{m\pm1} \cos(m\varphi_m-\theta_{m\pm1}),
  \nonumber \\
  s_{m\pm1} & = & K_{m\pm1} \sin(m\varphi_m-\theta_{m\pm1}),
\end{eqnarray}
so that after substitution into equation~\eqref{eq:relABandCph} we
obtain
\begin{equation}
  \label{eq:relphaseshifts}
  m \, \varphi_m - \theta_{m\pm1} = (m\pm1) \, \psi_{m\pm1},
\end{equation}
where $\varphi_m = \phi_m - \phi_\mathrm{los} + \pi/2$. This links the
phase-shift $\phi_m(R)$ of the $m$-spiral perturbation in the
gravitational potential with the phase-shifts $\psi_{m\pm1}(R)$ of the
$(m\pm1)$-spirals in the line-of-sight velocity field. The
corresponding pitch angles are related as
\begin{equation}
  \label{eq:psimvsphim}
  m \, \cot\zeta_m = (m\pm1) \, \cot\zeta_{m\pm1} 
  - \rmd \theta_{m\pm1}/ \rmd \ln R.
\end{equation}
In general, $K_{m\pm1}$ and $\theta_{m\pm1}$ depend in a rather
complex way on the gravitational potential, but as we show in
Appendix~\ref{app:gasorbits} they possess some generic properties. 

First, the amplitude $K_{m+1}$ is larger (smaller) than the amplitude
$K_{m-1}$ outside (inside) the corotation radius $\RCR$, and equal to
it at $\RCR$.  This implies a transition in the line-of-sight velocity
field at $\RCR$, going from the $(m+1)$-spiral dominating outside
$\RCR$ to the $(m-1)$-spiral dominating inside $\RCR$ (see also
Figure~\ref{fig:model_pl}). This is also concluded by
\cite{Schoenmakers1997} for the collisionless case and earlier by
\cite{Canzian1993} for the less general case of a tightly wound spiral
in the linear density-wave theory \citep[see also][]{Canzian1997}.
However, the dissipational nature of the gas, which we model in
Appendix~A via radial damping \citep[cf.][]{Wada1994}, can alter the
relative amplitudes of the harmonic terms (see also
Section~\ref{sec:spiralmodel} below).

Second, $\theta_{m\pm1}$ typically varies much less with radius than
the spiral phase-shifts, so that $\rmd \theta_{m\pm1} / \rmd \ln R$ in
equation~\eqref{eq:psimvsphim} is relatively small. As a result, we
can estimate the pitch angle $\zeta_m$ of the $m$-spiral perturbation
in the gravitational potential from the pitch angles $\zeta_{m-1}$
and/or $\zeta_{m+1}$ of the $(m-1)$-spiral and $(m+1)$-spiral in the
observed line-of-sight velocity field, without constructing a full
dynamical model (see also the last two panels of
Figure~\ref{fig:model_pl}).


\subsection{Radial flow velocity}
\label{sec:radialflowvelocity}

Nuclear spirals in principle provide a mechanism to transport gas from
kpc scales, where it often stalls inside a nuclear ring, into the
center of the galaxy. Still, as mentioned in Section~\ref{sec:intro},
if a nuclear spiral (or a nuclear bar) is due to gas moving on closed
elliptic orbits this results in non-circular motions, also referred to
as elliptic streaming, but not necessarily in net inflow towards
and/or outflow away from the center. However, unlike stars, gas is not
collisionless and its orbits interact and ex-change angular momentum
leading to net radial flows \citep[e.g.][]{Wada1994}.

Henceforth, in the analytic models in Appendix~\ref{app:gasorbits}, we
assume a weak perturbation in the gravitational potential causing gas
to deviate from circular onto elliptic orbits, while taking into
account its dissipative nature via radial damping. These analytic
spiral models are an extension of the analytic bar models introduced
by \citep{Wada1994}, who showed that they describe well the gas
behavior seen in hydrodynamical simulations. The radial damping causes
the gas to lose/gain angular momentum inside/outside the corotation
radius, which nicely matches the angular momentum transfer due to the
torque from the bar potential in numerical simulations. The amount of
radial damping, controlled through the dimensionless
parameter\footnote{Note that our $\lambda$ is the same as the
  dimensionless parameter $\Lambda$ in \cite{Wada1994}, while we
  define $\Lambda = 2 \lambda \kappa m (\Omega-\Omega_p)$.},
$\lambda$, thus provides a handle on the amount of net radial flow
that is needed to explain the observed non-circular motions in
addition to elliptic streaming.

In this way, a rather straightforward estimate of the net radial flow
velocity, $v_\mathrm{flow}$, can be obtained by comparing the radial
velocity $v_R$ (equation~\ref{eq:appsolvR}) of the analytic model that
includes radial damping ($\lambda>0$) with the analytic model without
radial damping ($\lambda=0$). At a given radius $R$, the maximum
radial velocity is given by
\begin{equation}
  \label{eq:vRmax}
  v_{R,\mathrm{max}} = m \, (\Omega-\Omega_p) \, R 
  \left( \frac{A^2+B^2}{\Delta^2+\Lambda^2} \right)^{1/2}.
\end{equation}
Here, $A$ and $B$ only depend on the weak perturbation as given in
equation~\eqref{eq:appdefAandB}, while
\begin{equation}
  \label{eq:defDeltaLambda}
  \Delta = \kappa^2 - m^2(\Omega-\Omega_p)^2,
  \quad
  \Lambda = 2 \lambda \kappa m (\Omega-\Omega_p),
\end{equation}
are functions of the angular frequency $\Omega(R)$ and the epicycle
frequency $\kappa(R)$ of the axisymmetric gravitational potential, as
well as the harmonic number $m$ and pattern speed $\Omega_p$ of the
weak perturbation.
Since without radial damping ($\Lambda=0$) all radial motion is due to
elliptic streaming, we subscribe a fraction
$|\Delta|/(\Delta^2+\Lambda^2)^{1/2}$ of the radial velocity to
elliptic streaming, leaving as an estimate of the radial flow velocity
\begin{equation}
  \label{eq:maxradflowvel}
  v_\mathrm{flow} = 
  \left( \frac{\Lambda^2}{\Delta^2+\Lambda^2} \right)^{1/2}
  v_R.
\end{equation}
Even though the analytic models neglect possible non-linear effects,
they can capture most features of observed non-circular motions (as we
show next in the case of NGC\,1097), and at the same time provide an
estimate of the fraction of the observed non-circular motions that is
due to net radial flow in addition to elliptic streaming.
Note that radial flow here does not mean that the gas is following
pure radial orbits with zero angular momentum, which would contribute
only to the harmonic term $s_1$ \citep[e.g.][]{Wong2004}.
Instead, the gas is expected to gradually spiral inward/outward as it
has both azimuthal and radial velocity components. In case the angular
momentum loss/gain is driven by a weak gravitational potential
perturbation with harmonic number $m$, this results in a contribution
to $c_{m\pm1}$ and $s_{m\pm1}$.

\section{Nuclear spiral in NGC\,1097}
\label{sec:ngc1097}

\begin{figure*}
  \begin{center}
    \includegraphics[width=1.0\textwidth]{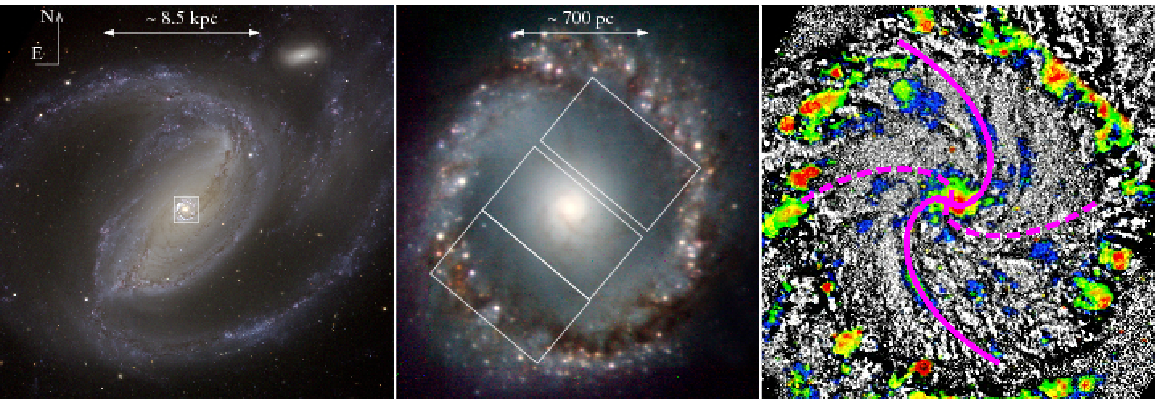}
  \end{center}
  \caption{\emph{Left}: \vlt/VIMOS color composite image of NGC\,1097,
    showing the large-scale spiral arms and bar with prominent dust
    lanes reaching down to the circumnuclear ring (credit: European
    Southern Observatory). \emph{Middle}: \vlt/NACO adaptive optics
    color-composite image of the circumnuclear ring region (credit:
    European Southern Observatory), with the footprints of the
    observations with the \gifu\ spectrograph. \emph{Right}: \hst/ACS
    structure map (20\arcsec$\times$20\arcsec) of the same
    circumnuclear region with the wavelet map of \cite{Lou2001}
    overplotted in color, with increasing intensity from blue to
    red. The solid magenta curves show the two-arm nuclear spiral with
    pitch angle $52$\dgr, which we predict based on the three-arm
    spiral structure in the velocity field. The dashed magenta curves
    indicate the two additional arms in case a $m=4$ spiral
    perturbation with the same pitch angle would be present (see
    Section~\ref{sec:disctwoarmspiral} for further details).}
  \label{fig:images}
\end{figure*}

We apply the above harmonic analysis to the observed emission-line
velocity field within the circumnuclear star forming ring of NGC\,1097.
We recover in the non-circular motions a spiral structure and infer
its pitch angle directly from the harmonic components. Next, we use a
spiral perturbation model to constrain the radial inflow velocity and
combine this with the gas density in the nuclear spiral to estimate
the mass inflow rate as function of distance from the center of
NGC\,1097.

\subsection{Non-circular motions}
\label{sec:noncircularmotions}

\begin{figure*}
  \begin{center}
    \includegraphics[width=1.0\textwidth]{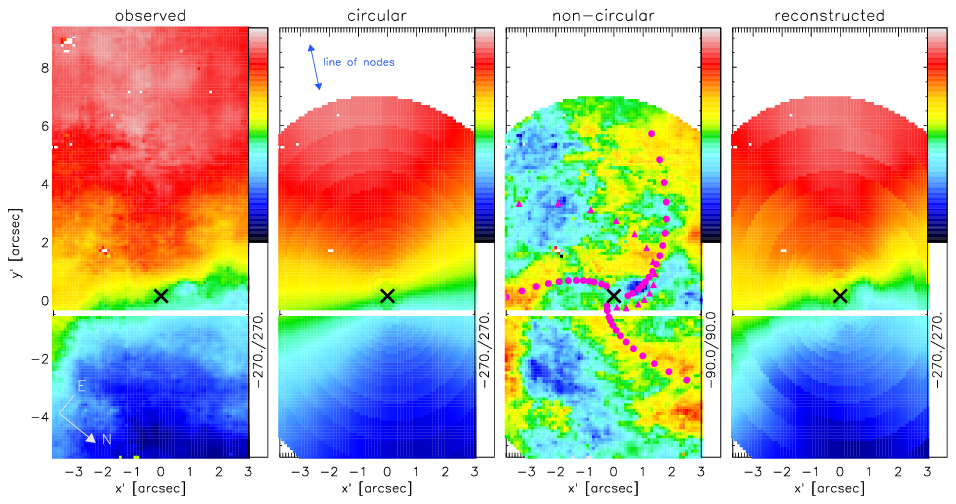}
  \end{center}
  \caption{The first panel shows the line-of-sight \NII\ emission-line
    velocity field within the circumnuclear ring in NGC\,1097,
    obtained with the \gifu\ spectrograph. The other panels show the
    results of applying the harmonic expansion of
    Section~\ref{sec:expansion}: the best-fit circular motion, the
    remaining non-circular motion, and the harmonic reconstruction
    from the sum of the latter two, taking into account the different
    (linear) velocity scale. The latter is indicated by the color bar
    at the right-hand side of each map, with the limits given below
    each bar. In the third panel, the magenta circles illustrate
    the three-arm logarithmic spiral with pitch angle $63$\dgr,
    derived from the third harmonic coefficients (see
    Fig.~\ref{fig:hardec_profiles} below). This is consistent with a
    weak two-arm spiral perturbation in the gravitational potential,
    which at the same time would also give rise to a one-arm
    logarithmic spiral with pitch angle $33$\dgr, indicated by the
    magenta triangles. Since both logarithmic spirals contribute in
    a different way it is not surprising that neither accurately
    traces the spiral structure in the non-circular motions (see
    Section~\ref{sec:noncircularmotions} for further details).}
  \label{fig:hardec_maps}
\end{figure*}

\begin{figure*}
  \begin{center}
    \includegraphics[width=1.0\textwidth]{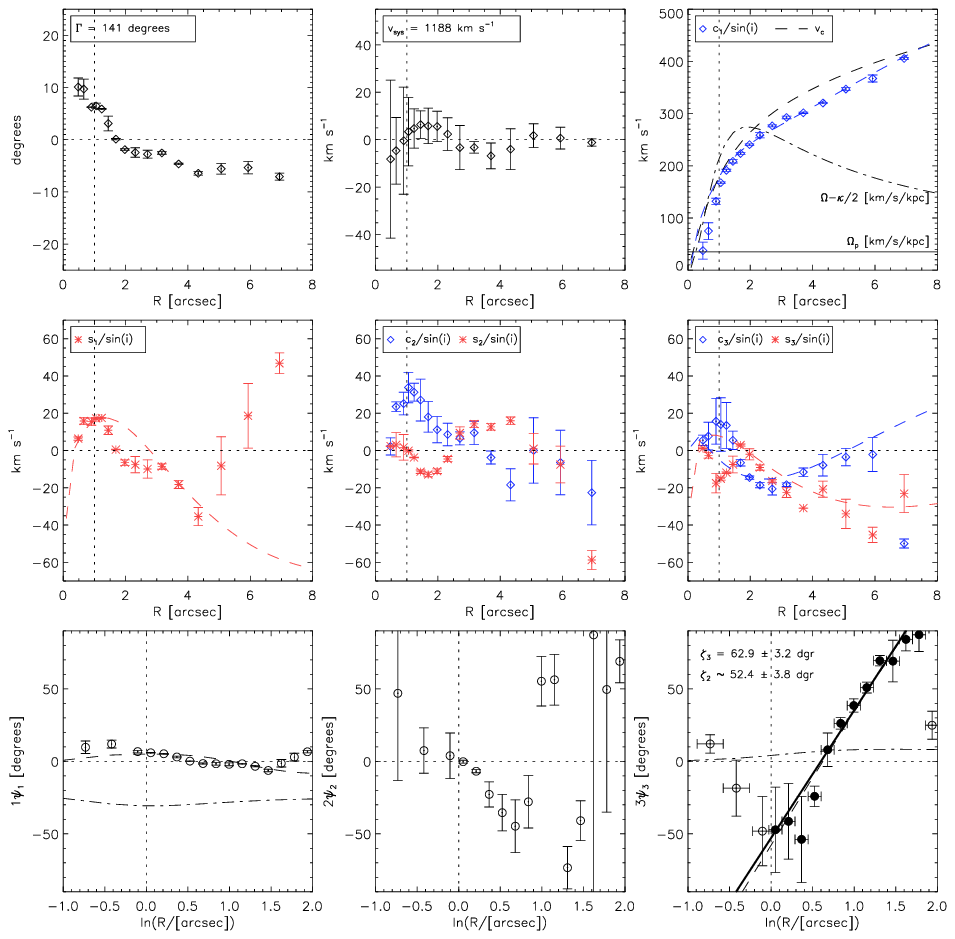}
  \end{center}
  \caption{Harmonic expansion of the observed line-of-sight velocity
    field of NGC\,1097. 
    After fixing the position angle $\Gamma$ and systemic velocity
    $V_\mathrm{sys}$ to their mean values indicated by the dotted line
    in, respectively, the \emph{top-left and top-middle panels}, the
    harmonic coefficients $c_m$ and $s_m$ are extracted as functions
    of radius $R$ (in arcsec). The coefficients have been divided by
    $q=\sin i$ to correct for the inclination $i=35$\dgr. The dotted
    vertical line in each panel indicates the radius $R=1\arcsec$,
    within which the measurements are less accurate due to possible
    contamination from the nuclear source and presence of significant
    amounts of dust.
    In the \emph{bottom panels}, the phase-shifts $\psi_m$ defined in
    equation~\eqref{eq:relABandCph} are plotted against $\ln R$, so
    that a spiral structure shows up as a smooth variation, which
    becomes linear in case of a logarithmic spiral as shown in
    equation~\eqref{eq:logmspiral}. 
    In this way, the \emph{bottom-right panel} shows
    that the structure in the non-circular motions is consistent with
    a three-arm logarithmic spiral with indicated pitch angle $\zeta_3$
    from the slope of the thick solid linear fit. 
    %
    The indicated estimate of the pitch angle $\zeta_2$ of the
    inferred two-arm spiral perturbation in the gravitational
    potential follows from equation~\eqref{eq:psimvsphim}, if we
    neglect the radial variation in the additional term
    $\theta_3$. 
    The latter is indeed expected to be small as indicated by the
    dot-dashed curve, which shows $\theta_3$ based on an analytical two-arm spiral
    perturbation model. Taking into account this term yields the
    model prediction for $\psi_3$ indicated by the dashed curve, which
    is nearly
    indistinguishable from the fitted thick solid line.
    Similarly, in the \emph{bottom-left panel}, the dot-dashed and dashed
    curves show the model prediction of $\theta_1$ and $\psi_1$, respectively.
    In line with the measurements, $\psi_1$ is small
    because the $c_1$ coefficient in $\tan\psi_1 = s_1/c_1$
    includes the dominating circular velocity $v_c$.
    The latter is shown as a black dashed curve in the\emph{ top-right
    panel}, whereas including the non-circular motions due to the
    modeled two-arm spiral perturbation, yields $c_1$ indicated by the blue
    dashed curve. Similarly, the model prediction for $s_1$ is given by the
    red dashed curve in the \emph{middle-left panel}, and those for $c_3$ and $c_3$
    by the blue and red dashed curves in the \emph{middle-right panel}.
    (See Section~\ref{sec:spiralmodel} and
    Appendix~\ref{app:gasorbits} for further details on the spiral
    perturbation model.)
  }
  \label{fig:hardec_profiles}
\end{figure*}

NGC\,1097 (ESO\,325--58) is a nearby (distance 14.5\,Mpc, so $1\arcsec
\simeq 70$\,pc) LINER/Seyfert\,1 host with a strong, $\simeq 16$\,kpc
long, bar and a $\simeq 0.7$\,kpc in radius circumnuclear star forming
ring (Figure~\ref{fig:images}). In this Sb galaxy dust can be traced
within the large-scale spiral arms out to a (outer Lindblad resonance,
OLR) radius of $\simeq 14$\,kpc, in prominent lanes along the bar, and
continuing within the nuclear ring as a spiral structure down to $\la
3.5$\,pc from the center \citep{Lou2001, Prieto2005}.
Non-circular motions associated with this nuclear spiral structure
\citep{Fathi2006, Davies2009}, indicate a possible mechanism to drive
gas from kpc scales down to a few pc from the center, where a
double-peaked broad H$\alpha$ emission profile
\citep{StorchiBergmann1993} indicates the presence of a super-massive
BH.

\cite{Fathi2006} describe in detail the observations and reduction of
the two-dimensional spectroscopy of NGC\,1097 obtained with the \gifu\
on the Gemini South Telescope (GS-2004B-Q-25, PI: Storchi-Bergmann).
Three pointings within the nuclear ring region
(Figure~\ref{fig:images}) provided 1500 individual spectra covering
5600--7000\,\AA\ at a velocity resolution of 85\,\kms\ and with a
spatial sampling of 0\farcs1.

In the first panel of Figure~\ref{fig:hardec_maps}, we present the \NII\
emission-line velocity field covering the inner $0.5 \times 1.0$\,kpc.
A systemic velocity of $v_\mathrm{sys} = 1188$\,\kms\ (equivalent to a
constant $c_0$ term) has been subtracted. For an inclination
$i=35$\dgr\ \citep[flattening $q=0.82$;][]{Fathi2006}, and position
angle of $\Gamma = 141$\dgr, the other panels show the results of
applying the harmonic expansion of Section~\ref{sec:expansion}.  From left
to right: the best-fit circular motion $V_\mathrm{circ} = c_1 \cos
\psi$, the remaining non-circular motion, and the harmonic
reconstruction from the sum of the latter two.

In Figure~\ref{fig:hardec_profiles}, the first two panels show the
adopted position angle and systemic velocity. The next four panels show
the coefficients $c_m$ and $s_m$ (in \kms) for the first three
harmonic terms, as function of the (deprojected) radius $R$. The
bottom three panels, show the phase shifts $\psi_m$ (in degrees) for
the alternative formulation of the harmonic expansion given in
equation~\eqref{eq:harmdefCph}, as functions of the (natural)
logarithm of $R$.

We interpret the non-circular residual motion (third panel of
Figure~\ref{fig:hardec_maps}) as a three-arm spiral structure. This
interpretation is supported by the significant amplitudes of $c_3$ and
$s_3$, and in particular by the the smooth variation of the
corresponding phase-shift $\psi_3$ with radius. Excluding the
uncertain measurements within $R=1\arcsec$ (vertical dotted line) and
the single measurement at the edge of the map, the remaining
measurements indicated by solid circles show a linear relation
between $3\psi_3$ and $\ln R$. This is consistent with a logarithmic
spiral defined in equation~\eqref{eq:logmspiral}, with fitted $a_0 =
1.82 \pm 0.33$\arcsec\ and $b_0 = -1.95 \pm 0.27$. The slope provides
a direct and robust measurement of the pitch angle of $\zeta_3 = 63
\pm 3$\dgr.

In the third panel of Figure~\ref{fig:hardec_maps}, this three-arm
logarithmic spiral with pitch angle $63$\dgr\ is plotted with
magenta circles on top of the non-circular motions.
If this three-arm spiral in the velocity field is due to a two-arm
spiral perturbation in the gravitational potential as we argue below,
it should also give rise to a one-arm logarithmic spiral in the
velocity field.
Combining equation~\eqref{eq:logmspiral} and the right-hand-side of
equation~\eqref{eq:relphaseshifts}, we find $\cot\zeta_1 =
\cot\zeta_3$ (and equal $a_0$).
This means a pitch angle $33$\dgr\ for the one-arm logarithmic spiral,
resulting in the magenta triangles in the third panel of
Figure~\ref{fig:hardec_maps}.
The contribution of the one-arm spiral to the non-circular motions is
partly absorbed into the circular motions since the $c_1$ coefficient
is indistinguishable from the circular velocity contribution. 
Moreover, as can be seen from Fig.~\ref{fig:hardec_profiles}, the
remaining $s_1$ coefficient contributes in a different way than the
combined $s_3$ and $c_3$ terms, so that it is not surprising that
neither the one-arm nor the three-arm spiral alone traces the spiral
structure in the non-circular motions.
This is even aside from possible additional contributions, or
contaminations in this case, from other (even) harmonic terms, as
discussed below in Section~\ref{sec:addnoncircmotion}.

The positive sign of the pitch angle indicates the nuclear spiral is
curved anti-clockwise with increasing radius, equivalent to the spiral
arms extending outwards of the large-scale bar, as can be seen in the
left panel of Figure~\ref{fig:images}. Since the global rotation in
NGC\,1097 is clockwise from the observed velocity field, and the
North-East is the ``far side'' from being more obscured, it follows
that both the large-scale and nuclear spirals are trailing.

\subsection{Two-arm spiral perturbation}
\label{sec:spiralmodel}

The above three-arm spiral structure in the non-circular motions is
consistent with a perturbation in the gravitational potential due to a
$m=2$ harmonic spiral (see also
Section~\ref{sec:potentialpertubation}).  From
equation~\eqref{eq:psimvsphim} it then follows that the pitch angle
$\zeta_2$ of this two-arm spiral perturbation in the gravitational
potential follows from the measured pitch angle $\zeta_3$ of the
three-arm spiral in the velocity field as $2\cot\zeta_2 = 3\cot\zeta_3
- \rmd \theta_3/\rmd \ln R$. We show below that, as expected, the
later term is small, so that for a measured $\zeta_3 \simeq 62 \pm
3$\dgr, we calculate $\zeta_2 \simeq 52 \pm 4$\dgr.

As illustrated in Figure~\ref{fig:model_pl} of
Appendix~\ref{app:gasorbits}, the true value of $\zeta_2$ might differ
slightly depending on the details of the gravitational potential
perturbation. We construct a model for the nuclear spiral in NGC\,1097
to get a handle on the latter difference as well as to constrain the
fraction of the observed non-circular motions that is due to net
radial inflow in addition to elliptical streaming.
We use the analytic solutions of Appendix~\ref{app:gasorbits} for
gaseous orbits in an axisymmetric gravitational potential,
$\Phi_0(R)$, that is weakly perturbed by a logarithmic $m=2$ spiral
with pitch angle $\zeta_2$.
These analytic spiral models are based on linearized equations of
motion under the epicycle approximation, but are not restricted to
tightly wound spirals \citep[e.g.][]{Lin1969, Canzian1997}.
In this way, we show below that the observed loosely wound nuclear
spiral in NGC\,1097 is still consistent with a density wave, and not
necessarily driven by shocks as suggested by \cite{Davies2009}.
Note that these simple models assume the existence of a weak
perturbation in the gravitational potential without specifying neither
how the perturbation arises nor how it is maintained. To include
potential driving mechanism such as the large-scale bar requires more
sophisticated models, which is beyond the goals and scope of this
paper.

We adopt the power-law model \citep{Evans1994} with axisymmetric
gravitational potential
\begin{equation}
  \label{eq:PLpot}
  \Phi_0(R) = 
  \begin{cases}
    v_0^2 \, \frac{2^{\beta/2}}{\beta} \left[ 1 -
      \left(1+\frac{R^2}{R_c^2}\right)^{-\beta/2} \right] &
    \text{$\beta \ne 0$}, \\
    v_0^2 \, \frac{1}{2} \left[ 1 - \ln\left(1+\frac{R^2}{R_c^2}\right)
    \right] & \text{$\beta = 0$},
  \end{cases}
\end{equation}
and corresponding circular velocity
\begin{equation}
  \label{eq:PLvcirc}
  v_c(R) = v_0 \, 2^{\beta/4} \frac{R}{R_c}
  \left(1+\frac{R^2}{R_c^2}\right)^{-(1/2+\beta/4)},
\end{equation}
so that $v_c(R_c)=v_0/2$ at the core radius $R_c$. The parameter
$\beta$ controls the logarithmic gradient of the rotation curve at
large radii: $\beta < 0$ rising, $\beta=0$ flat, and $\beta > 0$
falling. The three gravitational potential parameters $v_0$, $R_c$ and
$\beta$ are set by comparing the corresponding circular velocity
$v_c(R)$ with the measured radial profile of $c_1$, taking into
account the non-circular contribution due to the $m=2$ spiral
perturbation.
In the third panel of Figure~\ref{fig:hardec_profiles}, the black
dashed curve shows $v_c(R)$ for $v_0 = 275$\,\kms, $R_c = 1.2$\arcsec,
and $\beta=-0.6$. Except for the three values within $R=1$\arcsec,
this simple power-law model is a good representation of the $c_1$
measurements, once the non-circular contribution derived below is
added, as indicated by the blue dashed curve. The advantage of such a
simple analytic representation of the gravitational potential is that
it makes all subsequent calculations concerning the perturbation very
convenient.

For the amplitude of the gravitational potential perturbation we
assume $\Phi_2(R) = \epsilon_p \Phi_0(R)$, with constant strength
$\epsilon_p$, while the phase-shift is given by $\phi_2(R) =
\cot\zeta_2 \ln(R/a_0)$. Here, $a_0 = 1.8$\arcsec\ from the above fit
to the $m'=3$ harmonic terms in the line-of-sight velocity field,
while the corresponding approximation $\zeta_2 = 52$\dgr, is taken as
the initial value for the pitch angle. The three additional free
parameters are the pattern speed, $\Omega_p$, of the spiral
perturbation, the azimuthal viewing angle, $\phi_\mathrm{los}$, and
finally the amount of radial damping, $\lambda$.
Equations~\eqref{eq:appsolcmpm1} and~\eqref{eq:appsolsmpm1} then
provide predictions for the non-circular motion contribution in terms
of the harmonic coefficients $c_1$, $s_1$, $c_3$, and $s_3$, which we
compare with the corresponding measured radial profiles for NGC\,1097
in Figure~\ref{fig:hardec_profiles}.

The flattening of the \HI\ rotation curve in the outer parts of
NGC\,1097 \citep{Sofue1999} implies a nearly constant circular
velocity $v_c \simeq 300$\,\kms. With the angular and epicycle
frequencies approximately given by $\Omega \simeq \kappa/\sqrt{2}
\simeq v_c/R$, the OLR being at $R_\mathrm{OLR} \simeq 14$\,kpc yields
a pattern speed $\Omega_p \simeq \Omega + \kappa/2 \simeq
35$\,\kmskpc.  The latter places the corotation radius, $R_\mathrm{CR}
\simeq v_c/\Omega_p \simeq 8.6$\,kpc, or $\sim 10$\% beyond the extent
of the large-scale bar, consistent with numerical simulations of
``fast bars'' \citep[e.g.][]{Athanassoula1992, Debattista2000}, and
measured pattern speeds in similar galaxies
\citep[e.g.][]{Aguerri2003, Gerssen2003, Rautiainen2008}.
Assuming that the two-arm nuclear spiral as a gas density wave is
being driven by the large-scale bar \citep{Englmaier2000}, we adopt
the same value for the pattern speed of the perturbation. Since we
consider the harmonic coefficients well within the corotation radius,
they are not sensitive to, and hence do not constrain $\Omega_p$.

On the contrary, a small change in the azimuthal viewing angle already
causes a significant radial offset in the harmonic coefficients, so
that we need $\phi_\mathrm{los} \simeq (1.05 \pm 0.05) \, \pi/2$.
Next, too little radial damping results in too small amplitudes of the
coefficients $c_3$ and $s_3$ with respect to $s_1$ and the
non-circular contribution to $c_1$. Specifically, since $s_1$ and
$s_3$ are of similar amplitude and shape over most of the radial
range, significant radial damping with $\lambda>1$ is needed. As
described in Section~\ref{sec:radialflowvelocity} and discussed below
in Section~\ref{sec:massinflowrate}, this implies that net radial flow
makes up most of the intrinsic radial velocity.
Matching the amplitudes of the harmonics terms yields a strength of
the gravitational potential perturbation of $\epsilon_p \simeq 0.15$.
Finally, values for the pitch angle of the spiral perturbation that
are in the range of $\zeta_2 \simeq 52 \pm 4$\dgr, approximated above
from the pitch angle $\zeta_3$ of the three-arm spiral in the velocity
field, yield predictions for the harmonic coefficients that are
consistent with the measured harmonic coefficients.
The effect of the additional term $\rmd \theta_3/\rmd \ln R$ is indeed
small, so that $2\cot\zeta_2 \simeq 3\cot\zeta_3$ provides a robust
measurement of $\zeta_2$.

This is further illustrated in Figure~\ref{fig:hardec_profiles}, where
in addition to the measured harmonic coefficients, we show with dashed
curves the predictions of the above two-arm spiral perturbation model
with pitch angle $\zeta_2 = 52$\dgr, azimuthal viewing angle
$\phi_\mathrm{los} = 94.5$\dgr, and radial damping parameter $\lambda
= 2$.
Both $\theta_1$ and $\theta_2$, shown as dot-dashed curves in
respectively the bottom-left and bottom-right panel, indeed vary only
little with radius.
Similarly, the dashed curve in the bottom-right panel shows the
predicted relation between $3\psi_3$ and $\ln R$ while taking into
account $\theta_3$, which is nearly indistinguishable from the fitted
solid line.
The predicted harmonic terms match well the $c_3$ and $s_3$
measurements in the middle-right panel, and rather well the $s_1$
measurements in the middle-left panel, except for those at larger
radii. However, in particular the last three points in radius are less
certain because the corresponding ellipses to extract these
measurements are not fully covered (see also second panel of
Fig.~\ref{fig:hardec_maps}), and they might well be disturbed by the
nuclear ring.
The blue dashed curve in the top-right panel is the spiral model
prediction for $c_1$, which apart from the innermost measurements,
nicely traces the measured rotation curve.

Due to the significant non-circular motion contribution the latter is
different from the circular velocity of the (power-law) axisymmetric
gravitational potential shown as the black dashed curve.
The corresponding $\Omega(R)-\kappa(R)/2$ black dot-dashed curve is
still well above the pattern speed $\Omega_p = 35$\,\kmskpc\ (solid
horizontal line) at radius $\sim 8$\arcsec, which places the ILR well
beyond the nuclear ring radius of $\simeq 0.7$\,kpc.
If indeed the radius of nuclear rings is set by the location of the
ILR \citep[e.g.][]{Buta1996}, this indicates that the nuclear ring in
NGC\,1097 has migrated inward. Inward migration has also been
suggested for the nuclear ring in NGC\,4314 \citep{Benedict2002}, has
been seen in hydrodynamic simulations
\citep[e.g.][]{Fukuda2000,Regan2003}, and might be a consequence of
shepherding of the gas ring by star clusters that formed in it
\citep{vandeVen2009}.

\subsection{Mass inflow rate}
\label{sec:massinflowrate}

\begin{figure*}
  \begin{center}
    \includegraphics[width=0.8\textwidth]{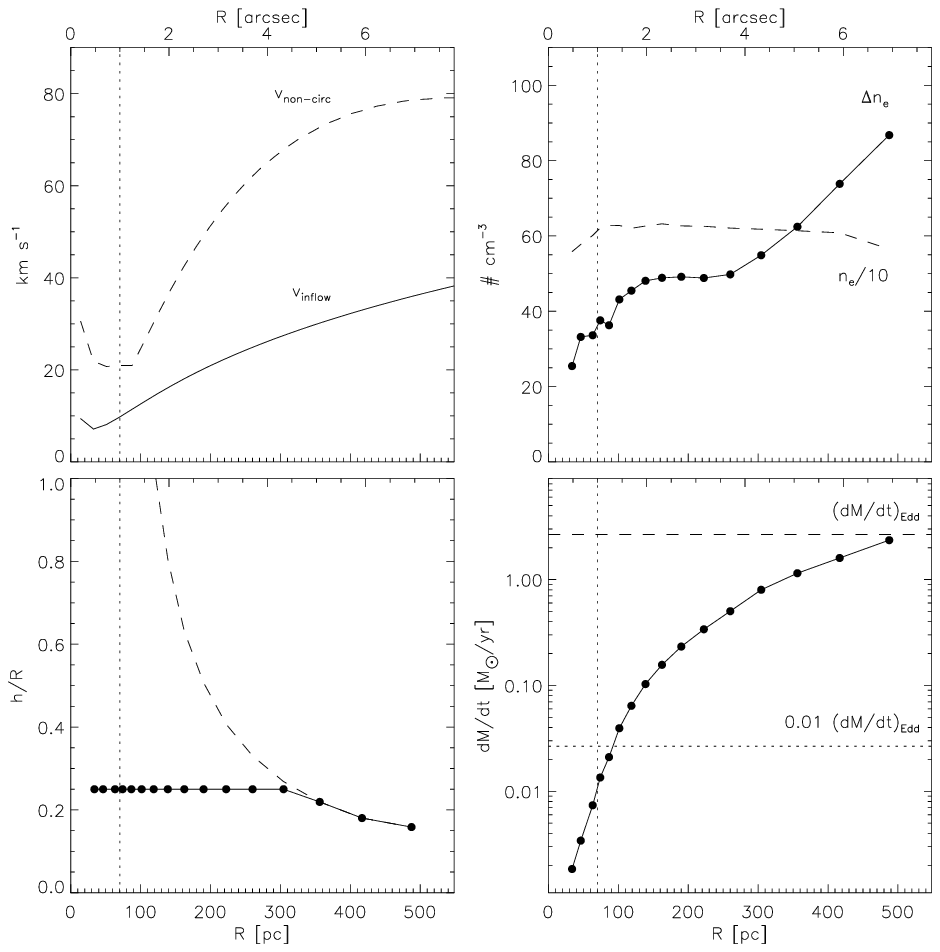}
  \end{center}
  \caption{Estimate of the mass inflow rate as function of distance
    $R$ (in pc) from the center or NGC\,1097. 
    \emph{Top-left:} The solid curve shows the radial inflow velocity
    inferred from a spiral perturbation model matched to the harmonic
    expansion of the velocity field. The radial inflow velocity is
    only a fraction of the non-circular motions indicated by the
    dashed curve.  
    \emph{Top-right:} The solid circles show the electron density
    $\dne$ in the nuclear spiral arms estimated from the variation in
    the flux ratio of the \SIIdoublet\ emission-line doublet. This
    (over)density is typically a factor ten smaller than the average
    electron density $\avne$ as indicated by the dashed line.
    \emph{Bottom-left:} The solid circles show the adopted scale
    height $h$, assuming a marginally stable disk resulting in the
    dashed curve, but with the relative width $R/(2m)$ of the $m=2$ spiral
    model in the equatorial plane as an upper limit.
    \emph{Bottom-right:} Combining the solid-curve values of the first
    three panels into equation~\eqref{eq:massinflowrate} results in the
    shown mass inflow rate (in \Msunyr). The upper dashed horizontal
    line indicates the Eddington accretion rate onto the central black
    hole in NGC\,1097 with a mass $\Mbh = 1.2 \times 10^8$\,\Msun\ 
    based on the central stellar velocity dispersion.  The lower
    dotted horizontal line at a factor $0.01$ of the Eddington
    accretion rate indicates the approximate transition from a
    Seyfert1 to a LINER active galactic nucleus, and is thought to be
    the rate below which mass accretion becomes radiatively
    inefficient.
    The dotted vertical line in each panel indicates the radius $R=1\arcsec$
    ($\simeq 70$\,pc) down to which the measurements are reliable.
    (See Sections~\ref{sec:massinflowrate} and~\ref{sec:feedingbh} for
    further details.)}
  \label{fig:mass_inflow_rate}
\end{figure*}

We found that a spiral perturbation model with only elliptic
streaming, in which gas is moving on closed elliptic orbits without
radial damping, cannot explain the observed non-circular motions in
NGC\,1097. In Section~\ref{sec:radialflowvelocity}, we showed that the
amount of radial damping required provides an estimate of the radial
flow velocity $v_\mathrm{flow}$ in terms of the radial velocity $v_R$
of the spiral model which matches the observed harmonic terms. Since
in the case of NGC\,1097 a large amount of radial damping $\lambda>1$
is needed, the ratio $v_\mathrm{flow}/v_R \simeq
|\Lambda|/\sqrt{\Delta^2+\Lambda^2}$ in
equation~\eqref{eq:maxradflowvel} is close to unity. In other words,
the contribution to the radial velocity in the nuclear spiral is
predominantly due to net radial flow, which is directed inward to the
center of NGC\,1097, as we are well within the corotation radius.
For the same spiral model as shown in Figure~\ref{fig:hardec_profiles}
with the dashed curves, the corresponding radial inflow velocity
$v_\mathrm{inflow}$ is shown in the top-left panel of
Figure~\ref{fig:mass_inflow_rate} with a solid curve. Even though nearly
identical to the radial velocity $v_R$, it still is only a fraction of
the total non-circular motion (dashed curve), which also includes the
azimuthal velocity.

To turn this radial inflow velocity into a constraint on the mass
inflow rate, we furthermore need to know the gas density in the
nuclear spiral, as well as the geometry or the area through which the
gas is flowing.
The flux ratio of the \SIIdoublet\ emission-line doublet included in
the spectral range of the \gifu\ observations allows us to constrain
the mean electron density $\avne$ over the observed field. 
Taking all values together yields an average flux ratio of $1.003 \pm
0.031$, which, adopting the prescription by \cite{Shaw1994} and using
the atomic parameters compiled by \cite{Mendoz1983} and
\cite{Osterbrock1989}, corresponds to $\avne = 600 \pm 77$\,\ccm\
assuming a mean electron temperature of $10^4$\,K.
Even though the flux ratio map might indicate a spiral structure
similar to that in the non-circular motions, the signal-to-noise is
not high enough to quantify the gas (over)density in the spiral arms
from it.  Instead, we apply the same elliptic annuli used to extract
the harmonic coefficients to the flux ratio map. From the distribution
within each annulus, we derive $\avne$ as a function of radius,
resulting in the dashed curve in the top-right panel of
Figure~\ref{fig:mass_inflow_rate} (divided by 10 for illustrative
purposes). Next, we assume that the width of the distribution,
indicated by the solid curve, is driven by the density wave contrast,
and hence yields an estimate
of the electron (over)density $\dne$ within the spiral arms.
Assuming a $50$\% lower (higher) electron temperate of $0.5 (1.5)
\times 10^4$\,K, leads to an increase (decrease) in the electron
densities by a factor of about 20\%.
%
%
In all cases, the density contrast is of the order of $10$\%, in
agreement with hydrodynamic simulations \citep[e.g.][]{Englmaier2000,
  Maciejewski2004b} and $K$-band imaging presented by
\cite{Davies2009}.
%
%
Using the proton mass and a factor 1.36 to account for the presence of
Helium, we convert $\dne$ to a gas mass overdensity $\drgas$.

For the geometry we assume in agreement with the spiral perturbation
model that both the radial flow velocity and gas mass overdensity, at
each radius $R$ in a disk with scale height $h$, vary as a sinusoidal
function in azimuthal angle $\psi$. Integrating over the $\psi$ values
for which the radial flow velocity is positive, the mass inflow rate
then reduces to
\begin{equation}
  \label{eq:massinflowrate}
  \Mdot = m \, v_\mathrm{inflow} \, \drgas \, \pi R^2 
  \, \frac{h}{R} \, \frac{1}{4m},
\end{equation}
for gas of density $\drgas$ flowing in the $m$ spiral arms towards the
center at a velocity  $v_\mathrm{inflow}$.
To estimate the scale height $h$, we start from a marginally stable
disk with Toomre's (\citeyear{Toomre1964}) $Q \simeq 1$, and substitute
in $Q = c_s \kappa/\pi G \Sigma_\mathrm{gas}$ a constant sound speed
$c_s \simeq 10$\,\kms, the epicycle frequency $\kappa$ of the
power-law axisymmetric gravitational potential, and
$\Sigma_\mathrm{gas} \simeq \rho_\mathrm{gas} h$, with
$\rho_\mathrm{gas}$ inferred from the average electron density
$\avne$. The dashed curve in the bottom-left panel of
Figure~\ref{fig:mass_inflow_rate} shows $h/R$ as function of radius.
Clearly, towards the center this leads to an unrealistically high
scale height, so that we constrain $h/R$ to be not larger than the
relative width $1/(2m)$ of a spiral arm, resulting in the solid curve.
Finally, in the bottom-right panel of Figure~\ref{fig:mass_inflow_rate},
we present the mass inflow rate $\dot{M}$ (in \Msunyr) as function of
radius.

\section{Discussion}
\label{sec:discussion}

We have argued for a two-arm spiral perturbation in the gravitational
potential as the source of the three-arm spiral structure in the
velocity field of NGC\,1097. We verify the implied two-arm spiral
distortion in the surface brightness, and discuss possible causes for
contributions from additional harmonic terms. Finally, we link the
mass inflow rate to the accretion onto the central BH.

\subsection{Two-arm spiral perturbation?}
\label{sec:disctwoarmspiral}

An $m=2$ spiral perturbation in the gravitational potential that would
explain the three-arm spiral structure in the velocity field of
NGC\,1097, implies an $m=2$ spiral distortion in the surface mass
density and hence a two-arm spiral structure in the surface
brightness.

Even if the latter distortion is too weak to measure directly, it
might show itself through obscuration by correlated dust.
Indeed, structure maps of NGC\,1097 \citep[e.g.][]{Pogge2002,
  Martini2003, Fathi2006} show spiral-like features, but they are not
evidently a two-arm nuclear spiral.
The structure map in the right panel of Figure~\ref{fig:images} is
based on the Richardson-Lucy image restoration technique
\citep{Snyder1993}, using a multi-step convolution of the \hst/ACS
high resolution camera FR656N image with a two-dimensional PSF model
constructed using Tiny Tim \citep{Krist1997}.
We have overplotted the structure map with the wavelet map from
\cite[][their Fig.~1]{Lou2001}, which does seem to be consistent with
a two-arm spiral structure, but additional spiral features cannot be
ruled out.

The solid magenta curves show the two-arm spiral with pitch angle
$\zeta_2 = 52$\dgr, which we inferred from the logarithmic spiral
fitted to the $m=3$ harmonic terms
(Section~\ref{sec:spiralmodel}). The two open arms trace well the
spiral structures in both the wavelet and structure map, except closer
to the nuclear ring when, at least in the northern part, a more
tightly wound spiral seems needed.
The dashed magenta curves indicate the two additional arms in case of
an $m=4$ spiral perturbation with the same pitch angle. Such
higher-order \emph{even} harmonic terms can result from non-linear
coupling of modes, and ---although smaller in amplitude then the $m=2$
spiral perturbation--- might give rise to possibly additional spiral
features.
Furthermore, a spiral perturbation driven by the large-scale bar is
not necessarily restricted to an $m=2$ harmonic term\footnote{For
  example, the axisymmetric power-law potential in
  equation~\eqref{eq:PLpot}, being perturbed by replacing the radius $R$ by
  $R^2=x^2+(y/q)^2$ with $q<1$, creates besides $m=2$ also
  higher-order even harmonic terms.}. 
Still, as long as the perturbation is bi-symmetric all resulting
harmonic terms are even.

%
Nonetheless, \cite{Prieto2005} note in high-resolution \vlt/NACO
infrared images a central spiral with a three-arm symmetry, though one
of the three arms does not seem to continue towards the nuclear ring,
but instead splits into a number of spiral filaments.
\cite{Davies2009} show the inner $4\arcsec \times 4\arcsec$ of the
NACO $J$-band residual image from \cite{Prieto2005} together with
their SINFONI $K$-band residual image (their Fig.~1).
While the third arm is already weaker in the $J$-band (in particular
when taking into account the narrow intensity scaling that is
saturating the two strong arms), it nearly disappears in the $K$-band.
Also, the residual flux distribution of (warm) H$_2$, as traced by the
2.12\,$\mu$m 1--0\,S(1) line (their Fig.~3), reveals two strong arms.
Nevertheless, \cite{Davies2009} claim a (weak) third arm in the
stellar and gas density to give rise to the \emph{two}-arm spiral
structure they argue to see in the residual H$_2$ velocity field
(their Fig.~5).
However, when the inner $2\arcsec$ of our \NII\ non-circular motions
(third panel of Fig.~\ref{fig:hardec_maps}) are overlayed by their
H$_2$ residual velocity field (as in their Fig.~6), we believe the
(observed and predicted) \emph{three}-arm spiral structure is traced
inwards by the H$_2$ kinematics.
This apparent agreement is not obvious as their near-infrared
observations are less affected by dust than our optical measurements,
but also because it is not evident that (warm) H$_2$ emission and
ionized emission trace the gas kinematics in the same way.  
Even so, we find that the data presented by \cite{Davies2009} do not
contradict our interpretation.

As indicated by \cite{Davies2009}, the kinematics of the stars do not
show any significant deviations from axisymmetry (see their
Fig.~2). Indeed, it is expected that the bulge stars are dominating
the stellar kinematics, and that the perturbation is only visible in
the intensity due to dust extinction in the equatorial plane.
Since the stability of the gas in the equatorial plane inhibits
self-amplification \citep[see also][]{Davies2009}, the observed spiral
pattern in the gas is most likely due to a weak spiral perturbation in
the total, stellar-dominated, gravitational potential.
The most natural driver of this perturbation is the large-scale bar,
which might induce spiral shocks in the gas
\citep[e.g.][]{Maciejewski2004a, Maciejewski2004b}. 
\cite{Davies2009} argue for these shocks to be present in the inner
region of NGC\,1097, based on a large amplitude of the radial motions
with respect to the velocity dispersion of the gas.
However, as discussed above (see also top-left panel of Fig.~4), only
part of the observed non-circular motions might be radial motions, and
shocks might just be the trigger to create long-lived gas density
waves \citep{Englmaier2000, Ann2005}.
Whether spiral shocks and/or bulge stars ionize the perturbed gas, the
spiral perturbation can be traced through non-circular motions in the
observed ionized gas kinematics.

In this way, we find a three-arm structure in the \NII\ non-circular
motions, which we believe is consistent with the residual H$_2$
velocity field, as well as with two spiral arms visible in the
residual H$_2$ flux, $K$-band image, and $J$-band image, presented by
\cite{Davies2009}.
All this leads to our interpretation of a weak \emph{two}-arm spiral
perturbation in the gravitational potential driven by the large-scale
bar; though further modeling is needed to find out if for example the
spiral pattern in the gas are induced by shocks or are long-lived
density waves.
Instead, the explanation by \cite{Davies2009} of a \emph{three}-arm
spiral perturbation in the gravitational potential, requires quite a
special driving mechanism, such as non-linear interactions between the
large-scale bar and a nuclear bar or a (dark) massive orbiting compact
object, which so far have not been demonstrated in models.
In addition, observed structure in addition to that expected from a
two-arm spiral perturbation in the gravitational potential, including
a possible weak third arm, might be the result of asymmetric dust
obscuration, as we discuss next.

\subsection{Additional non-circular motion?}
\label{sec:addnoncircmotion}

The coefficients $c_m$ and $s_m$, that provide the phase-shift
$\psi_m(R)$, may be ``contaminated'' by additional contributions to
the non-circular motion, but in general they are not expected to
result in a smooth variation of the phase-shift with radius, as seen
for a spiral structure.
An exception is the $m=1$ harmonic term, since the measured
coefficient $c_1$ also incorporates the circular velocity as $v_c\sin
i$, which in general dominates over the non-circular motions
$c_1^\mathrm{nc}$ and $s_1$. This results in a phase-shift $\psi_1$
that is everywhere close to zero, and due to the degeneracy in $c_1$
there is little hope of constraining a two-arm spiral perturbation
from its contribution to the $m=1$ harmonic term in the velocity
field.
However, we can predict its pitch angle $\zeta_1$ and phase-shift
$\phi_1^\mathrm{nc}$ from the (logarithmic) spiral inferred from the
contribution to the $m=3$ harmonic term: $\tan\zeta_1 = 3\tan\zeta_3$
and $\psi_1^\mathrm{nc}(R) = 3\psi_1^\mathrm{nc}(R)$. As a result, we
might use $c_1^\mathrm{nc} = s_1 \cot \psi_1^\mathrm{nc}$ as an
estimate of the non-circular contribution to $c_1$.
The effect is shown in the top-right panel of
Figure~\ref{fig:hardec_profiles}, where the black dashed curve is
$v_c$ of the (power-law) axisymmetric potential, which after taking
into account $c_1^\mathrm{nc}$ yields the blue dashed curve that
matches the measured $c_1$ coefficients indicated by the blue
diamonds.
This provides a novel way to correct for non-circular motions, which
otherwise might, for example, lead to an underestimation of the
(inner) slope of the mass distribution \citep[e.g.][]{Hayashi2006}.

Figure~\ref{fig:hardec_profiles} shows that, besides $m=1$ and $m=3$
harmonic terms in the velocity field expected from a $m=2$
perturbation in the gravitational potential, the non-circular motions
also seem to contain an $m=2$ harmonic term. There are several effects
that might (partly) cause this additional contribution.

As shown by \citeauthor{Schoenmakers1997}
(\citeyear{Schoenmakers1997}, their equation~7), an error in the
(kinematic) center results in a spurious $m=0$ and $m=2$ contribution
to the line-of-sight velocity as
\begin{eqnarray}
  \label{eq:dvlos_dxdy}
  \delta V_\mathrm{los} & = & \Vs
  \Bigl[ \Bigr. (1+\alpha) \frac{\delta x'}{2 R} 
  \nonumber \\
  & & - (1-\alpha) \Bigl( 
    \frac{\delta x'}{2 R} \cos 2\psi
    + \frac{\delta y'}{2 R} \sin 2\psi
    \Bigr)
  \Bigl. \Bigr],
\end{eqnarray}
We see that the effect on $c_2$ and $s_2$ vanishes if $\alpha = 1$,
i.e., when the circular velocity curve increases linearly with radius,
$v_c \propto R$. As in most galaxies, the latter is also the case in
the inner region of NGC\,1097, but still $c_2$ is significantly
positive within the central $\lesssim 2\arcsec$. Not surprisingly,
varying the kinematic center (in the process of finding the best-fit
set of ellipses as described in Section~\ref{sec:expansion}) does not
remove the $m=2$ harmonic contribution, but in contrast makes the
overall fit worse. This makes a significant effect due to an incorrect
center unlikely.

Based on the analysis in Section~\ref{sec:potentialpertubation}, we expect
a similar contribution of both $m=0$ and $m=2$ harmonic terms to the
velocity field from an $m=1$ distortion of an axisymmetric
distribution.
It is unlikely that the gravitational potential itself is lopsided
since the stars that dominate in mass do not show any such signature.
Also, the nuclear spiral itself is expected to be bi-symmetric if it is
indeed a gas density wave driven by the large-scale bar.
Still, the phase-shift $2\psi_2$ in the bottom-middle panel of
Figure~\ref{fig:hardec_profiles} seems to vary smoothly, and even close
to linearly as a function of $\ln R$ within the central $\lesssim
2\arcsec$ where prominent dust features are present.
Moreover, the amplitude of the slope is similar to the linear relation
of $3\psi_3$ versus $\ln R$ in the bottom-right panel, but with
negative instead of positive sign. This is consistent with a
logarithmic spiral with the same pitch angle but orientated clockwise
instead of anti-clockwise, i.e., leading instead of trailing.
In principle, both leading and trailing nuclear spirals can exist, as
shown by \cite{Wada1994}, but in their simulations the smaller leading
spiral dissolves before the long-lived larger trailing spiral fully
emerges.
Alternatively, the spiral-like contribution to the $m=2$ harmonic
term, appearing like a ``negative image'', might result from an
asymmetric dust obscuration, mimicking a lopsided distortion.

The often prominent dust lanes along the leading edges of bars in
galaxies are associated with shocks in the gas streaming along the
length of the bar \citep[e.g.][]{Athanassoula1992}, which in turn lead
to velocity jumps across the dust lanes \citep[e.g.][]{Mundell1999}.
Moreover, numerical models of dust lanes \citep[e.g.][]{Gerssen2007}
as well as analytical models of diffuse disks
\citep[e.g.][]{Valotto2004} show that dust extinction can have a
significant effect on the velocity along the line-of-sight.
%
%
Henceforth, we expect the dust and possible shocks associated with the
spiral features to distort the velocity field, but the modeling
required to understand the specific effects on the non-circular motion
is beyond the scope of this paper.

\subsection{Feeding the central black hole?}
\label{sec:feedingbh}

The Eddington accretion rate onto a central black hole
\begin{equation}
  \label{eq:MdotEdd}
  \MEdd = 2.2 \, \mathrm{M}_\odot \mathrm{yr}^{-1} \;
  \left( \frac{\epsilon}{0.1} \right)^{-1} 
  \left( \frac{\Mbh}{10^8 \mathrm{M}_\odot}\right),
\end{equation}
adopting $\epsilon=0.1$ for the radiative efficiency, and a mass $\Mbh
= 1.2 \times 10^8$\,\Msun\ for the central black hole in NGC\,1097
---based on the measured central stellar velocity dispersion of
$\sigma_\star = 196 \pm 5$\,\kms\ \citep{Lewis2006} substituted in the
$\Mbh-\sigma_\star$ relation \citep{Tremaine2002}--- yields $\MEdd
\simeq 2.7$\,\Msunyr. This value is indicated in the bottom-right
panel of Figure~\ref{fig:mass_inflow_rate} by the upper dashed
horizontal line.
The lower dotted horizontal line at $\Mdot = 0.01 \, \MEdd$ is the
approximate transition from a Seyfert1 to a LINER active galactic
nucleus \citep[see for a review][]{Ho2005}, and thought to be the rate
below which mass accretion becomes radiatively inefficient \citep[see
for reviews][]{Quataert2001, Narayan2005}.

Whereas NGC\,1097 is typically classified as a LINER galaxy,
monitoring of the nucleus reveals evolution in its activity up into
the Seyfert1 regime \citep{StorchiBergmann2003}. \cite{Nemmen2006}
find that the observed optical to X-ray spectral energy distribution
in the nucleus of NGC\,1097 is consistent with an inner radiatively
inefficient accretion flow plus outer standard thin disk, with a mass
\emph{accretion} rate of $\Mdot \simeq 6.4 \times 10^{-3} \,\MEdd$.
We find a mass \emph{inflow} rate, down to $\Mdot \simeq
0.011$\,\Msunyr ($\simeq 4.2 \times 10^{-3}\,\MEdd$) at a distance
$R=1\arcsec$ ($\simeq 70$\,pc) from the center, where the gas
kinematics are still accurately measured and well described by the
two-arm spiral perturbation model.
These constraints are obtained at very different scales (tenths versus
tens of pc), and the mass inflow rate and the onset of nuclear
activity are not necessarily linked in time. Even so, it is
encouraging that we obtain comparable values.

Our mass inflow rate is significantly lower than $\Mdot \sim
0.6$\,\Msunyr\ estimated by \cite{StorchiBergmann2007a} at a distance
$R=100$\,pc.
In the latter estimate, a non-circular motion of $50$\,\kms\ is
adopted for the inflow velocity, whereas we find from our spiral model
that even though the contribution of the radial inflow dominates over
elliptic streaming, it is still only a fraction ($\simeq 13$\,\kms\ at
$100$\,pc) of the non-circular motions.
Moreover, the estimated electron density of $\sim 500$\,\ccm\ is
similar to the average density we derive directly from emission line
ratios, but more than an order of magnitude higher than the expected
(over)density in the spiral arms ($\simeq 43$\,\ccm\ at $100$\,pc).
Finally, instead of assuming a fixed opening angle of $20$\dgr, we
incorporate the geometry from the spiral model, and allow for the
scale height to vary, in order for the disk to remain marginally
stable.
In this way, we obtain an estimate of the mass inflow rate of similar
value as the expected mass accretion rate onto the central BH in
NGC\,1097, without having to invoke an unclear filling factor
\citep[e.g.][]{StorchiBergmann2007b}.
Even so, the mass inflow rate might be even further constrained, in
particular by measuring emission line ratios at higher signal-to-noise
to obtain a more accurate estimate of the (over)density, as well as
using other emission lines to break the degeneracy with temperature.

\section{Summary and conclusions}
\label{sec:summaryandconcl}

We presented harmonic expansion of the line-of-sight velocity field as
a suitable method to identify and quantify possible structures in the
non-circular motions, including nuclear spirals.
We confirmed earlier findings \citep{Canzian1993, Schoenmakers1997}
that a weak perturbation in the gravitational potential of harmonic
number $m$, causes the surface brightness to also exhibit an $m$
distortion, but leads to $m-1$ and $m+1$ harmonic terms in the
velocity field.
In the case of a $m$-arm spiral perturbation in the gravitational
potential, we found that the corresponding $(m+1)$-arm and $(m-1)$-arm
spirals in the velocity field are respectively less and more tightly
wound, with pitch angles approximately related as $m \cot\zeta_m
\simeq (m\pm1)\cot\zeta_{m\pm1}$.
In Appendix~\ref{app:gasorbits}, we derived an analytic perturbation
model, which allows for a simple estimate of the fraction of the
measured non-circular motions that is due to radial flow.

We applied this method to the emission-line velocity field within the
circumnuclear star forming ring of NGC\,1097, obtained with the \gifu\ 
spectrograph.
The radial variation of the resulting $m=3$ harmonic terms can be
fitted with a logarithmic spiral with a pitch angle $\zeta_3 = 63 \pm
3$\dgr.
We linked these $m=3$ harmonic terms in the velocity field to a weak
perturbation of the gravitational potential due to a two-arm nuclear
spiral with an inferred pitch angle $\zeta_2 \simeq 52 \pm 4$\dgr.
This predicts a two-arm spiral distortion in the surface brightness,
as hinted by the dust structure in central images of NGC~1097,
although additional spiral structure might be present as a result of
higher-order even harmonic terms.
Furthermore, this two-arm spiral perturbation of the gravitational
potential adds a combined $m=1$ and $m=3$ spiral structure to the
velocity field, as revealed in the non-circular motions of the ionised
gas in the center of this galaxy.
We argued that it is unlikely that the presence of also $m=2$ harmonic
terms in the velocity field is due to lopsidedness in the
gravitational potential nor in the nuclear spiral itself, which we
expect to be due to long-lived density waves in the gas driven by the
bi-symmetric large-scale bar.
Instead, we postulated that an asymmetric dust obscuration mimics a
lopsided distortion, and gives rise to the additional even harmonic
terms in the velocity field.

To match the measured odd harmonic terms in the velocity field of
NGC\,1097, a spiral perturbation model with a large amount of radial
damping is required. This indicates both strong dissipation and, that, in
addition to elliptic streaming, a significant fraction of the
non-circular motions is due to radial inflow.
We combined the inferred radial inflow velocity with the gas density
in, and the geometry of, the spiral arms to estimate the mass inflow
rate as function of radius.
We calculated the gas density from the variation in the flux ratio of
the \SIIdoublet\ emission-line doublet, and we incorporated in the
geometry the scale-height assuming a marginally stable disk.
The resulting mass inflow rate decreases to $\Mdot \simeq
0.011$\,\Msunyr\ at a distance of $R=70$\,pc from the center, down to
where our measurements are still reliable.
We showed that this corresponds to $\Mdot \simeq 4.2 \times
10^{-3}\,\MEdd$, where the latter Eddington accretion rate is onto a
black hole in NGC\,1097 with a mass $\Mbh = 1.2 \times 10^8$\,\Msun\ 
based on the central stellar velocity dispersion.
This mass inflow rate is consistent with the active galactic nucleus
in NGC\,1097 varying between LINER and Seyfert1, but comparison with
mass accretion models are hampered by the (still) very different
physical scales as well as possible time delay between mass inflow and
the onset of nuclear activity.

We conclude that the line-of-sight velocity can provide not only a
cleaner view on nuclear spirals than does the associated dust, but
that the presented method also allows one to quantitatively study
these possibly important links in fueling the centers of galaxies,
including a handle on the mass inflow rate as a function of radius.


\section*{Acknowledgments}
\label{sec:acknowledgments}

It is a pleasure to thank Eva Schinnerer for helpful suggestions, and
Paul Wiita for discussions and a critical reading of the manuscript.
We thank the referee, Eric Emsellem, for constructive comments on this
work.
GvdV acknowledges support provided by NASA through Hubble Fellowship
grant HST-HF-01202.01-A awarded by the Space Telescope Science
Institute, which is operated by the Association of Universities for
Research in Astronomy, Inc., for NASA, under contract NAS 5-26555.
KF is supported by the Swedish Research Council (Vetenskapsr{\aa}det),
and gratefully acknowledges the hospitality of the Institute for
Advanced Study, to which a visit contributed greatly in completing
this project.
In Figure~\ref{fig:images}, the \vlt/VIMOS images (left panel) have
been taken and pre-processed by European Southern Observatory (ESO)
Paranal Science Operation astronomers, with additional image
processing by Henri Boffin (ESO), and the \vlt/NACO images (middle
panel) is based on research published in the October issue of
Astronomical Journal, vol.\ 130, p.\ 1472.


\appendix

\section{Gaseous orbits in a weakly perturbed gravitational potential}
\label{app:gasorbits}

\begin{figure*}
  \begin{center}
    \includegraphics[width=1.0\columnwidth]{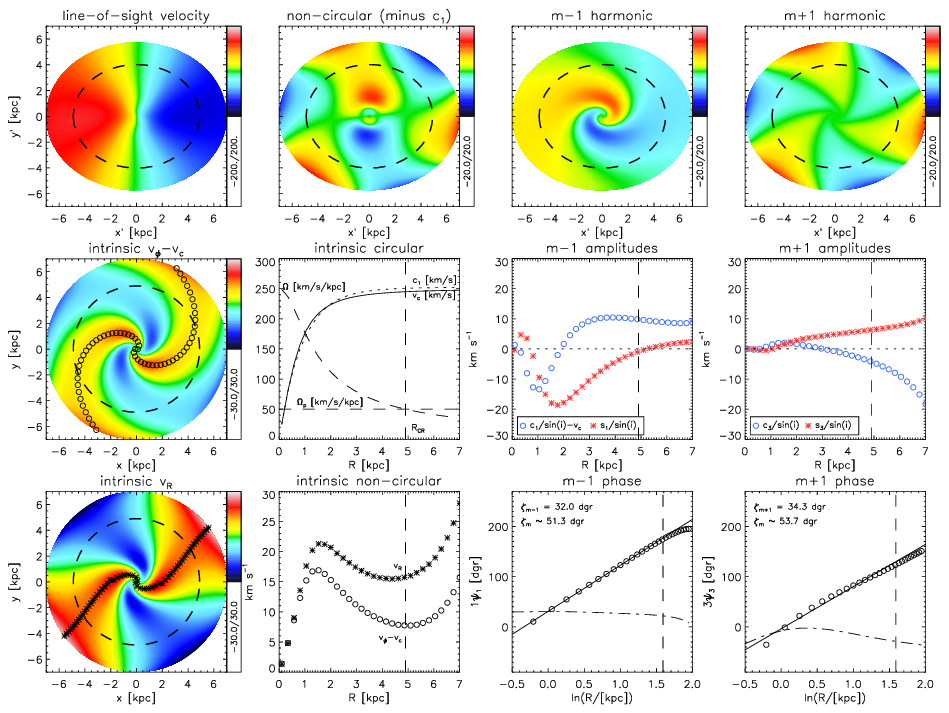}
  \end{center}
  \caption{An axisymmetric logarithmic potential perturbed by a weak
    $m=2$ logarithmic spiral with pitch angle $\zeta_2 = 50$\,\dgr.
    The top panels show for an inclination $i=30$\dgr,
    \emph{projected} maps on the sky-plane $(x',y')$ of the
    line-of-sight velocity, non-circular motions, and separately, the
    contribution of the $m-1$ and $m+1$ harmonic terms.
    The first panels in the middle and bottom rows show the
    \emph{intrinsic} maps in the equatorial plane $(x,y)$ of
    respectively the azimuthal and radial velocity due to the spiral
    perturbation. The overplotted open circles and stars trace the
    maxima as function of radius, resulting in the radial profiles
    shown in the second panel in the bottom row.
    The second panel in the middle row shows the radial profiles of
    the circular velocity $v_c$ of the axisymmetric potential,
    while the dotted curve shows $c_1$, which includes the
    contribution from the spiral perturbation. Moreover, the falling
    dashed curve is the angular frequency $\Omega$, which intersects
    the horizontal dashed line of the assumed pattern speed
    $\Omega_p$, at the corotation radius $R_\mathrm{CR}$. The latter is
    indicated in all panels either by a vertical dashed line or dashed
    ring on the maps.
    The third and fourth panels in the middle row show the radial
    profiles of the harmonic coefficients $c_{m\pm1}$ and $s_{m\pm1}$,
    divided by $\sin i$ to correct for inclination.
    The third and fourth panels in the bottom row show the
    corresponding phase shifts $\phi_{m\pm1}$, defined in
    equation~\eqref{eq:relABandCph}, against $\ln R$. In this way, a
    spiral structure shows up as a smooth variation, which becomes
    nearly linear because the assumed spiral perturbation is
    logarithmic as in equation~\eqref{eq:logmspiral}.
    The slopes of the fitted solid lines yield the indicated values of
    the pitch angles $\zeta_{m\pm1}$ of the $(m\pm1)$-arm spirals
    shown in the corresponding projected maps in the third and fourth
    panels in the top row.
    In turn, based on equation~\eqref{eq:psimvsphim}, both values
    provide an estimate as indicated of the pitch angle $\zeta_2$ of
    the $m=2$ spiral perturbation. The mild difference with the input
    value $\zeta_2 = 50$\,\dgr\ is due to the radial variation of the
    additional terms $\theta_{m\pm1}$ indicated by the dot-dashed
    lines.  
  }
  \label{fig:model_pl}
\end{figure*}

We derive analytic solutions of gaseous orbits in a weakly perturbed
gravitational potential \citep[for modeling of a strong perturbation
see e.g.][]{Spekkens2007}. We follow the treatment of
\citeauthor{BT87} (\citeyear{BT87}, p.\ 146-148) for a weak harmonic
perturbation. Like \cite{Schoenmakers1997}, we include an additional
phase-shift $\phi_m(R)$ as in equation~\eqref{eq:potharmm} to
accommodate for a spiral perturbation. Since gas at the same spatial
location has the same velocity, we look for closed loop orbits that do
not intersect themselves. However, to take into account the
dissipative nature of gas, we include, like \cite{Wada1994}, radial
damping, so that the first-order equation of motion becomes
\begin{equation}
  \label{eq:appdiffeqR}
  \ddot{R}_1 + 2\lambda\kappa_0\dot{R}_1 + \kappa_0^2 R_1 = 
  - R_0 \left( A \cos\eta + B \sin\eta \right),
  \qquad
  \eta = m[\phi_0(t)-\phi_m(R_0)],
\end{equation}
where $\lambda$ controls the amount of radial damping. The subscript
zero refers to zeroth-order with constant $R_0$ and
$\phi_0(t)=(\Omega_0-\Omega_p)t$, with $\Omega_p$ the pattern speed of
the perturbation. We have introduced the epicycle frequency, $\kappa^2
= 2 \Omega^2 (1+\alpha)$, and angular frequency, $\Omega = v_c/R$,
with $\alpha = \rmd\ln v_c / \rmd\ln R$ the logarithmic slope of the
circular velocity, $v_c^2(R) = \rmd \Phi_0(R) / \rmd \ln R$, of the
axisymmetric part of the gravitational potential.
Furthermore,
\begin{equation}
  \label{eq:appdefAandB}
  A = \frac{\Phi_m}{R^2} 
  \left[ \frac{2}{1-\omega} + \frac{\rmd\ln \Phi_m}{\rmd\ln R} \right],
  \quad
  B = \frac{\Phi_m}{R^2}
  m \cot\zeta_m,
\end{equation}
where we have introduced $\omega = \Omega_p/\Omega$, and $\zeta_m$ is
the pitch angle as defined in equation~\eqref{eq:pitchangle}.  
Solving equation~\eqref{eq:appdiffeqR} we find that the solution for
closed loop orbits is
\begin{eqnarray}
  \label{eq:appsolR}
  R & = & R_0 [1 -  (a \cos\eta + b \sin\eta) ],
  \\
  \label{eq:appsolphi}
  \phi & = & \phi_0 + [ 2 (a \sin\eta - b \cos\eta) - \xi \sin\eta ] / [m(1-\omega)]
  \\
  \label{eq:appsolvR}
  v_R & = & v_c \left\{ m (1-\omega) [a \sin\eta - b \cos\eta] \right\},
  \\
  \label{eq:appsolvphi}
  v_\phi & = & v_c \left\{ 1 + (1+\alpha) [ (a-\xi) \cos\eta + b \sin\eta) \right\}, 
\end{eqnarray}
with $\alpha$ appearing in the last line because of a first-order
conversion from guiding center $(R_0,\phi_0)$ to a point $(R,\phi)$ in
the observed velocity field. Moreover, 
\begin{equation}
  \label{eq:appdefabxi}
  a = \frac{A\Delta-B\Lambda}{\Delta^2 + \Lambda^2},
  \quad
  b = \frac{B\Delta+A\Lambda}{\Delta^2 + \Lambda^2},
  \quad
  \xi = \frac{1}{\kappa^2} \frac{\Phi_m}{R^2} \frac{2}{1-\omega},
\end{equation}
where we have defined $\Delta = \kappa^2 - m^2(\Omega-\Omega_p)^2$ and
$\Lambda = 2 \lambda \kappa m (\Omega-\Omega_p)$.
The orbit solutions have a singularity at the corotation radius where
$\Omega=\Omega_p$ ($\omega=1$), because the adopted epicycle
approximation breaks down. Without radial damping
($\lambda=\Lambda=0$), the (collisionless) orbit solutions also have
singularities when $\Delta=0$, i.e., at the Lindblad resonances given
by $\Omega-\kappa/m=\Omega_p$.

Next, we assume to first order $\phi_0 \approx \phi$ and replace it by
$\psi = \phi - \phi_\mathrm{los} + \pi/2$ which is zero along the line
of nodes. We also define $\varphi_m = \phi_m - \phi_\mathrm{los} +
\pi/2$, so that $\eta = m[\psi-\varphi_m(R)]$. This allows us to
recast the above expressions for $v_R$ and $v_\phi$ in multiple angles
of $\psi$ as in equations~\eqref{eq:solvR} and~\eqref{eq:solvphi}, with
\begin{eqnarray}
  \label{eq:appsolcR}
  c_R & = & -m(1-\omega) [a\sin(m\varphi_m) + b\cos(m\varphi_m)],
  \\
  \label{eq:appsolsR}
  s_R & = & m(1-\omega) [a\cos(m\varphi_m) - b\sin(m\varphi_m)],
  \\
  \label{eq:appsolcphi}
  c_\phi & = & (1+\alpha)(a - \xi) \cos(m\varphi_m) 
  - (1+\alpha) b \sin(m\varphi_m).
  \\
  \label{eq:appsolsphi}
  s_\phi & = & (1+\alpha)(a - \xi) \sin(m\varphi_m) 
  + (1+\alpha) b \cos(m\varphi_m),
\end{eqnarray}
Substitution in equation~\eqref{eq:vlosharm} results in the expression
for the line-of-sight velocity in equation~\eqref{eq:solvlos}, where
the harmonic coefficients are given by
\begin{eqnarray}
  \label{eq:appsolcmpm1}
  c_{m\pm1} & = & 
  A_\pm \cos(m\varphi_m) - B_\pm \sin(m\varphi_m)
  \\
  \label{eq:appsolsmpm1}
  s_{m\pm1} & = & 
  A_\pm \sin(m\varphi_m) + B_\pm \cos(m\varphi_m),
\end{eqnarray}
with
\begin{equation}
  \label{eq:appdefABpm}
  A_\pm = \frac12 \Vs \, \left\{ \, [(1+\alpha) \mp m(1-\omega)] \, a 
    - (1+\alpha) \xi \, \right\},
  \qquad
  B_\pm = \frac12 \Vs \, [(1+\alpha) \mp m(1-\omega)] \, b.
\end{equation}
Substituting $A_\pm = K_{m\pm1} \cos\theta_{m\pm1}$ and $B_\pm =
K_{m\pm1} \sin\theta_{m\pm1}$, with
\begin{equation}
  \label{eq:appdefKthpm}
  K_{m\pm1}^2 = A_\pm^2 + B_\pm^2,
  \qquad
  \tan\theta_{m\pm1} = B_\pm/A_\pm,
\end{equation}
we arrive at the form given in equation~\eqref{eq:harmcoeffphim}.

In Figure~\ref{fig:model_pl}, we present an example of a weakly
perturbed axisymmetric logarithmic potential defined in
equation~\eqref{eq:PLpot}, with $v_0 = 250$\,\kms, $R_c = 1.0$\,kpc
and $\beta=0$. The amplitude of the perturbation is a factor
$\epsilon_p=0.02$ times the axisymmetric logarithmic potential, while
the angular dependence is due to a logarithmic spiral defined in
equation~\eqref{eq:logmspiral}, with pitch angle $\zeta_2 = 50$\dgr.
This results in non-circular motions which contribute a 1-arm and
3-arm spiral structure to the observed line-of-sight velocity field as
shown in Figure~\ref{fig:model_pl} for an adopted inclination of
$i=30$\dgr.

In general, $K_{m\pm1}$ and $\theta_{m\pm1}$ in
equation~\eqref{eq:appdefKthpm} above depend in a rather complicated
way on the gravitational potential. However, in case of the linear
spiral density-wave theory \citep{Shu1973, Canzian1997} the
expressions reduce to
\begin{equation}
  \label{eq:denswave}
  K_{m\pm1} = \frac12 v_m \sin i \;  
  \left[ \frac{\kappa}{2\Omega} \mp 
    \frac{m(\Omega-\Omega_p)}{\kappa} \right],
  \qquad
  \tan\theta_{m\pm1} = \pm \cot\zeta_m,
\end{equation}
where the constant amplitude $v_m$ measures the strength of a \emph{tightly}
wound spiral without radial damping.
In this case, $\cot\zeta_m \gg 1$ and $\lambda=\Lambda=0$, so that $b
= B/\Delta \gg a = A/\Delta$, and $b \gg \xi$ in
equation~\eqref{eq:appdefabxi}, and hence $\tan\theta_{m\pm1} = B/A
\propto \cot\zeta_m$ and $K_{m\pm1} \simeq B_\pm$. The expression for
$B_\pm$ in equation~\eqref{eq:appdefABpm} is proportional to the
reduced expression for $K_{m\pm1}$ in equation~\eqref{eq:denswave},
since by substituting $\kappa^2=2\Omega^2(1+\alpha)$ and
$\omega=\Omega_p/\Omega$, the term in square brackets in
equation~\eqref{eq:denswave} is proportional to $[(1+\alpha) \mp
m(1-\omega)]$ in equation~\eqref{eq:appdefABpm}.




\end{document}